\documentclass[preprint,12pt,authoryear]{elsarticle}

\usepackage{amssymb}
\usepackage{amsmath}
\usepackage{graphicx}
\usepackage{subcaption}
\usepackage{placeins}
\usepackage{amsfonts}
\usepackage{booktabs}
\usepackage[colorlinks=true, allcolors=blue]{hyperref}
\usepackage[symbol]{footmisc}

\journal{Journal of Commodity Markets}

\begin{document}

\begin{frontmatter}



\title{Multi-Factor Function-on-Function Regression of Bond Yields on WTI Commodity Futures Term Structure Dynamics}


\author[inst1]{Peilun He\corref{cor1}}

\affiliation[inst1]{organization={Department of Actuarial Studies and Business Analytics, Macquarie University},
    city={Macquarie Park},
    citysep={}, 
    postcode={2109}, 
    state={NSW},
    country={Australia}}

\ead{peilun.he@mq.edu.au}

\cortext[cor1]{Corresponding author}

\author[inst2]{Gareth W. Peters}

\affiliation[inst2]{organization={Department of Statistics and Applied Probability, University of California Santa Barbara},
    city={Santa Barbara},
    citysep={}, 
    postcode={93106}, 
    state={CA},
    country={United States}}

\author[inst3]{Nino Kordzakhia}

\affiliation[inst3]{organization={School of Mathematical and Physical Sciences, Macquarie University},
    city={Macquarie Park},
    citysep={}, 
    postcode={2109}, 
    state={NSW},
    country={Australia}}

\author[inst1]{Pavel V. Shevchenko}

\begin{abstract}
In the analysis of commodity futures, it is commonly assumed that futures prices are driven by two latent factors: short-term fluctuations and long-term equilibrium price levels. In this study, we extend this framework by introducing a novel state-space functional regression model that incorporates yield curve dynamics. Our model offers a distinct advantage in capturing the interdependencies between commodity futures and the yield curve. Through a comprehensive empirical analysis of WTI crude oil futures, using US Treasury yields as a functional predictor, we demonstrate the superior accuracy of the functional regression model compared to the Schwartz-Smith two-factor model, particularly in estimating the short-end of the futures curve. Additionally, we conduct a stress testing analysis to examine the impact of both temporary and permanent shocks to US Treasury yields on futures price estimation.
\end{abstract}




\end{frontmatter}

\newpage


\section{Introduction}
\label{sec:intro}

The analysis of commodity futures plays a critical role for both practitioners and academics, serving purposes such as risk management, price forecasting, portfolio diversification, and supply chain optimisation. In real world markets, commodity futures are significantly influenced by economic conditions. Economic recessions, in particular, affect both the demand and supply dynamics of commodities, directly impacting prices. For instance, during a recession, economic activities slow down, reducing the need for energy commodities such as crude oil and natural gas.

The yield curve, often regarded as a barometer of global economic conditions, typically takes one of two forms: contango, where short-term interest rates are lower than long-term rates, or backwardation, where long-term rates are lower than short-term rates. Researchers have shown that a backwardation structure, also known as an inverted yield curve, is a reliable indicator of an impending economic recession \citep[see, e.g.,][]{estrella1996yield, haubrich1996predicting, ang2006does, wright2006yield, zaloom2009read, chinn2015predictive}. Investigating the relationship between commodity futures and yield curves is thus of considerable importance. For instance, \cite{hess2008commodity} demonstrated that inflation-related news significantly affects commodity futures prices during recessions. \cite{hu2013commodity} examined the predictive power of U.S. futures on East Asian economies, while \cite{rogel2014seasonal} studied the impact of financial crises on the seasonal component of natural gas prices.

Early studies on commodity futures focused on modelling futures prices using latent factors. \cite{gibson1990stochastic} applied the Ornstein-Uhlenbeck (OU) process to oil futures in a two-factor model, representing spot price and convenience yield. Building on this framework, \cite{schwartz2000short-term} modelled the logarithm of crude oil spot prices as the sum of two latent factors. These factors capture short-term fluctuations and the long-term equilibrium price level, respectively. This latent factor model and its extensions have since become widely adopted in commodity futures modelling.

Further developments in this modelling approach have enhanced its applicability. For example, \cite{eydeland1999fundamentals} introduced a multi-factor model for electricity markets, incorporating deterministic seasonality and additional stochastic components, modelled using Levy processes. Time-changed Levy processes have also become standard in option pricing, revealing jump behaviour and volatility dynamics \citep{carr2004time, huang2004specification, fallahgoul2023risk}. \cite{sorensen2002modeling} extended the model with three hidden factors, including a seasonal component, to capture agricultural commodity price dynamics, while \cite{kiesel2009two} focused directly on modelling electricity futures prices. \cite{ames2020risk} introduced time-varying drift and mean reversion speed parameters for crude oil futures. More recently, \cite{han2022correlated, han2024autoregressive} adapted the two-factor model for European Unit Allowances futures by assuming serially correlated measurement errors, and \cite{he2024multi} generalised the Schwartz-Smith model using a polynomial diffusion model to capture higher-order, non-linear dynamics. 
Moreover, \cite{cortazar2019commodity} integrated analysts’ forecasts to improve spot price estimations, and \cite{peters2013calibration}  developed a partial Markov Chain Monte Carlo method to manage non-linear, non-Gaussian multi-factor models. Comparative studies have also highlighted different models' capabilities. For instance, \cite{schwartz1997the} evaluated models with up to three factors, including convenience yield and interest rates, across copper, oil, and gold markets. Similarly, \cite{cortazar2006an} found that three- and four-factor models excel in explaining the term structure of crude oil futures, with the four-factor model particularly effective in capturing volatility dynamics.

However, this class of models is not without limitations. A common assumption is that futures prices depend on a few hidden factors, often specific to a particular market. In reality, commodity markets are interconnected and influenced by both local and global economic conditions. This paper addresses this limitation by extending the Schwartz-Smith two-factor model to incorporate the interdependencies between the crude oil futures market and the bond yields market. The proposed model is versatile and can be applied to investigate interdependencies between commodity futures markets and other financial markets.

This paper is structured as follows. Section \ref{sec:model} introduces the Schwartz-Smith two-factor model for pricing commodity futures, followed by its extension with a functional regression component to capture the interdependencies between yield curves and futures prices. Section \ref{sec:functional_representation} discusses how the functional regression is transformed into a finite sum of factors, extracted through kernel Principal Component Analysis (kPCA). In Section \ref{sec:estimation}, we employ the Kalman filter to jointly estimate the hidden state variables and the unknown model parameters. A comprehensive empirical analysis is presented in Section \ref{sec:empirical_analysis}, including a comparison between the traditional two-factor model and the proposed functional regression model. Additionally, stress testing is performed to assess the impact of shocks in bond yields on futures price estimates. Finally, Section \ref{sec:conclusion} concludes the paper.

\section{Two-Factor Functional Regression Model}
\label{sec:model}

In this section, we present the two-factor functional regression model. In Section \ref{sec:SS_model}, we introduce the widely used two-factor model for pricing commodity futures, initially proposed by \cite{schwartz2000short-term}. In Section \ref{sec:fr_model}, we extend this two-factor model by incorporating a functional regression component that captures the interdependencies between the futures curve and the yield curve.

\subsection{Schwartz-Smith Two-Factor Model}
\label{sec:SS_model}

We begin by introducing the two-factor model initially proposed by \cite{schwartz2000short-term}. This model and its extensions are widely applied in the modelling of commodity futures and other types of futures \citep[see, e.g.,][]{sorensen2002modeling, manoliu2002energy, casassus2005stochastic, cortazar2006an, favetto2010parameter, peters2013calibration, cortazar2019commodity, ames2020risk}.

We model the logarithm of spot price $S_t$, for $t \in \{1, \dots, N\}$, as the sum of two hidden factors $\chi_t$ and $\xi_t$: 
\begin{equation}
    \log{(S_t)} = \chi_t + \xi_t, 
    \label{eq:spot}
\end{equation}
where $\chi_t$ represents the short-term fluctuation and $\xi_t$ is the long-term equilibrium price level. We assume that both $\chi_t$ and $\xi_t$ follow an Ornstein–Uhlenbeck (OU) process:
\begin{equation}
d\chi_t = -\kappa_{\chi} \chi_t dt + \sigma_{\chi} d W_t^{\chi},
\label{eq:SS_chi}
\end{equation}
and 
\begin{equation}
d\xi_t = (\mu_{\xi} - \kappa_{\xi} \xi_t)dt + \sigma_{\xi} dW_t^{\xi},
\label{eq:SS_xi}
\end{equation}
where $\kappa_{\chi}, \kappa_{\xi} \in \mathbb{R}^+$ are the speed of mean-reversion parameters, $\mu_{\xi} \in \mathbb{R}$ is the mean level of the long-term factor, and $\sigma_{\chi}, \sigma_{\xi} \in \mathbb{R}^+$ are the volatility parameters. The processes $(W_t^{\chi})_{t \ge 0}$ and $(W_t^{\xi})_{t \ge 0}$ are correlated standard Brownian motions with the correlation coefficient $\rho$. In the original model by \cite{schwartz2000short-term}, only the short-term factor $\chi_t$ follows the OU process, while the long-term factor $\xi_t$ follows a geometric Brownian motion with $\kappa_{\xi} = 0$. However, in this paper, we allow $\kappa_{\xi}$ to take any positive value. A similar setup can be found in \cite{manoliu2002energy, casassus2005stochastic, peters2013calibration, ames2020risk}. For consistency, we continue to refer to this extended model, with $\kappa_{\xi} \ge 0$, as the Schwartz-Smith two-factor model, or simply the Schwartz-Smith (SS) model.

Assuming constant risk premiums $\lambda_{\chi}, \lambda_{\xi} \in \mathbb{R}$, the risk-neutral processes of $\chi_t$ and $\xi_t$ are given by: 
\begin{equation}
d\chi_t = (-\kappa_{\chi} \chi_t - \lambda_{\chi}) dt + \sigma_{\chi} d W_t^{\chi*}, 
\label{eq:SS_rn_chi}
\end{equation}
and 
\begin{equation}
d\xi_t = (\mu_{\xi} - \kappa_{\xi} \xi_t - \lambda_{\xi})dt + \sigma_{\xi} dW_t^{\xi*}, 
\label{eq:SS_rn_xi}
\end{equation}
where $W_t^{\chi*}$ and $W_t^{\xi*}$ are correlated standard Brownian motions under the risk-neutral measure. 

In discrete time, given the initial values $\chi_{0}$ and $\xi_{0}$, $\chi_t$ and $\xi_t$ are jointly normally distributed with mean 
$$\mathbb{E}^*\left( \left. \left[ \begin{matrix}
    \chi_t \\
    \xi_t
\end{matrix} \right] \right| \mathcal{F}_{0} \right) = 
\left[ \begin{matrix}
    e^{-\kappa_{\chi} t}\chi_{0} - \frac{\lambda_{\chi}}{\kappa_{\chi}} \left(1 - e^{-\kappa_{\chi} t}\right) \\
    e^{-\kappa_{\xi} t}\xi_{0} + \frac{\mu_{\xi} - \lambda_{\xi}}{\kappa_{\xi}} \left(1 - e^{-\kappa_{\xi} t}\right) 
\end{matrix} \right]$$
and covariance matrix
$$Cov^*\left( \left. \left[ \begin{matrix}
    \chi_t \\
    \xi_t
\end{matrix} \right] \right| \mathcal{F}_{0} \right) =  \left[\begin{matrix}
\frac{1 - e^{-2\kappa_{\chi} t}}{2\kappa_{\chi}} \sigma_{\chi}^2 & \frac{1 - e^{-(\kappa_{\chi} + \kappa_{\xi}) t}}{\kappa_{\chi} + \kappa_{\xi}} \sigma_{\chi} \sigma_{\xi} \rho \\
\frac{1 - e^{-(\kappa_{\chi} + \kappa_{\xi}) t}}{\kappa_{\chi} + \kappa_{\xi}}\sigma_{\chi}\sigma_{\xi}\rho & \frac{1 - e^{-2\kappa_{\xi} t}}{2\kappa_{\xi}}\sigma_{\xi}^2
\end{matrix}\right], $$
where $\mathcal{F}_t$ is the natural $\sigma$-algebra generated up to time $t$. The operators $\mathbb{E}^*(\cdot)$ and $Cov^*(\cdot)$ denote the expectation and covariance under the risk-neutral processes. Therefore, the spot price $S_t$, which is the sum of $\chi_t$ and $\xi_t$, follows a log-normal distribution with 
\begin{align}
\log[\mathbb{E}^*(S_t | \mathcal{F}_{0})] &= \mathbb{E}^*[\log(S_t) | \mathcal{F}_{0}] + \frac{1}{2}Var^*[\log(S_t) | \mathcal{F}_{0}] \nonumber\\
&= e^{-\kappa_{\chi} t}\chi_{0} + e^{-\kappa_{\xi} t}\xi_{0} + A(t), \nonumber
\end{align}	
where
\begin{align}
A(t) =& -\frac{\lambda_{\chi}}{\kappa_{\chi}}(1 - e^{-\kappa_{\chi} t}) + \frac{\mu_{\xi} - \lambda_{\xi}}{\kappa_{\xi}}(1 - e^{-\kappa_{\xi} t}) \nonumber \\
&+ \frac{1}{2}\left(\frac{1 - e^{-2\kappa_{\chi} t}}{2\kappa_{\chi}}\sigma_{\chi}^2 + \frac{1 - e^{-2\kappa_{\xi} t}}{2\kappa_{\xi}}\sigma_{\xi}^2 + 2\frac{1 - e^{-(\kappa_{\chi} + \kappa_{\xi})t}}{\kappa_{\chi} + \kappa_{\xi}}\sigma_{\chi}\sigma_{\xi}\rho \right).
\label{eq:At}
\end{align}
The function $A(t)$ depends solely on time $t$ and is independent of the latent factors $\chi_t$ and $\xi_t$. 

Let $F_{t, T}$ denote the futures price at time $t$ with maturity 
$T$. Under the arbitrage-free assumption, the futures price equals the expected spot price at maturity $T$, given all available information at time $t$. Therefore, under the risk-neutral measure, we have (assuming the interest rate is non-stochastic)
\begin{equation}
Y_t(T) = \log{(F_{t, T})} = \log{[\mathbb{E}^*(S_T | \mathcal{F}_t)]} = A(T-t) + e^{-\kappa_{\chi} (T-t)} \chi_t + e^{-\kappa_{\xi} (T-t)} \xi_t.
\label{eq:SS_futures}
\end{equation}
After discretising the real world processes \eqref{eq:SS_chi} and \eqref{eq:SS_xi}, the dynamics of the state vector $\boldsymbol{X}_t$ follow a vector autoregressive (VAR) process: 
\begin{equation}
\boldsymbol{X}_t = \boldsymbol{C} + \boldsymbol{E} \boldsymbol{X}_{t-1} + \boldsymbol{v}_t, 
\label{eq:SS_xt}
\end{equation}
where 
$$\boldsymbol{X}_t = \left[ \begin{matrix} \chi_t \\ \xi_t \end{matrix} \right],\; \boldsymbol{C} = \left[ \begin{matrix} 0 \\ \frac{\mu_{\xi}}{\kappa_{\xi}} \left(1 - e^{-\kappa_{\xi} \Delta t} \right) \end{matrix} \right],\; \boldsymbol{E} = \left[ \begin{matrix} e^{-\kappa_{\chi} \Delta t} & 0 \\ 0 & e^{-\kappa_{\xi} \Delta t}\end{matrix} \right], $$
and $\boldsymbol{v}_t$ is a vector of correlated normally distributed noise term with $\mathbb{E}(\boldsymbol{v}_t) = \textbf{0}$ and
$$Cov(\boldsymbol{v}_t) = \boldsymbol{\Sigma}_v = \left[\begin{matrix}
\frac{1 - e^{-2\kappa_{\chi} \Delta t}}{2\kappa_{\chi}} \sigma_{\chi}^2 & \frac{1 - e^{-(\kappa_{\chi} + \kappa_{\xi}) \Delta t}}{\kappa_{\chi} + \kappa_{\xi}}\sigma_{\chi}\sigma_{\xi}\rho \\
\frac{1 - e^{-(\kappa_{\chi} + \kappa_{\xi}) \Delta t}}{\kappa_{\chi} + \kappa_{\xi}}\sigma_{\chi}\sigma_{\xi}\rho & \frac{1 - e^{-2\kappa_{\xi} \Delta t}}{2\kappa_{\xi}}\sigma_{\xi}^2
\end{matrix}\right].$$
Here, $\Delta t$ is the time step between $(t-1)$ and $t$. To express the logarithm of the futures price as a linear measurement equation, we add a random noise term, resulting in:
\begin{equation}
\boldsymbol{Y}_t = \boldsymbol{D}_t + \boldsymbol{F}_t \boldsymbol{X}_t + \boldsymbol{w}_t, 
\label{eq:SS_yt}
\end{equation}
where 
$$\boldsymbol{Y}_t = \left( \log{(F_{t,T_1})}, \dots, \log{(F_{t,T_P})} \right)^\top, \boldsymbol{D}_t = \left( A(T_1 - t), \dots, A(T_P - t) \right)^\top, $$
$$\boldsymbol{F}_t = \left[ \begin{matrix} e^{-\kappa_{\chi} (T_1 - t)}, \dots, e^{-\kappa_{\chi} (T_P - t)} \\ e^{-\kappa_{\xi} (T_1 - t)}, \dots, e^{-\kappa_{\xi} (T_P - t)} \end{matrix} \right]^\top, $$
and $T_1 < T_2 < \dots < T_P$ denote the maturity time of each futures contract. The noise vector $\boldsymbol{w}_t$ is $P$-dimensional, normally distributed with $\mathbb{E}(\boldsymbol{w}_t) = \textbf{0}$ and
$$Cov(\boldsymbol{w}_t) = \boldsymbol{\Sigma_w} = \left[ \begin{matrix} \sigma_1^2 & 0 & \dots & 0 \\ 0 & \sigma_2^2 & \dots & 0 \\ \vdots & \vdots & \ddots & \vdots \\ 0 & 0 & \dots & \sigma_P^2 \end{matrix} \right]. $$
Equations \ref{eq:SS_xt} and \ref{eq:SS_yt} together form the complete Schwartz-Smith two-factor model.

\subsection{Functional Regression Representation}
\label{sec:fr_model}

In this section, we extend the Schwartz-Smith (SS) model by incorporating a functional regression component based on bond yields. This extension offers the advantage of bridging the commodity futures market and the bond yields market, enabling a deeper analysis of the impact of bond yields on futures prices. In recent years, functional data analysis (FDA) has gained prominence in various areas of economics and finance. For example, \citet{hyndman2009forecasting} proposed a weighted functional principal component regression method to forecast functional time series for the Australian fertility rate. \citet{hays2012functional} introduced a functional dynamic factor model for US Treasury yields, simultaneously estimating hidden factor time series and functional factor loadings. Additionally, \citet{horvath2020functional} developed functional methods for forecasting forward curves, demonstrating their superiority over multivariate methods.

We extend the SS model by adding a functional regression component to the futures price formulation in Equation \eqref{eq:SS_futures}: 
\begin{equation}
    Y_t(T_i) = A(T_i - t) + e^{-\kappa_{\chi} (T_i - t)} \chi_t + e^{-\kappa_{\xi} (T_i-t)} \xi_t + \int_{0}^{\tau_M} \gamma_i(s) Z_t(s) ds + w_t(T_i), 
    \label{eq:fr_futures}
\end{equation}
where $Z_t(\cdot)$ is the yield curve at time $t$, and $\gamma_i(\cdot)$ are the functional coefficients. In this study, we assume that the futures curve is influenced only by the yield curve at the same point in time, with no dependence on the past yield curves. Additionally, we assume that $\gamma_i(\cdot)$ is time-invariant. The parameter $\tau_M = T_M - t$ represents the time to maturity of the longest futures contract.

The state equations of the SS model remains unchanged:
\begin{equation}
    \chi_t = e^{-\kappa_{\chi} \Delta t} \chi_{t-1} + v_{t, \chi}
    \label{eq:fr_chi}
\end{equation}
and 
\begin{equation}
    \xi_t = \frac{\mu_{\xi}}{\kappa_{\xi}} \left(1 - e^{-\kappa_{\xi} \Delta t} \right) + e^{-\kappa_{\xi} \Delta t} \xi_{t-1} + v_{t, \xi}, 
    \label{eq:fr_xi}
\end{equation}
where $v_{t, \chi}$ and $ v_{t, \xi}$ are normally distributed noise term. We refer to Equations \eqref{eq:fr_futures}, \eqref{eq:fr_chi}, and \eqref{eq:fr_xi} collectively as the two-factor functional regression (FR) model. 

\section{Functional Representation and Transformation}
\label{sec:functional_representation}

To transform the integral representation of the functional regression into a vector operation suitable for linear estimation procedures, we introduce the kernel principal component analysis (kPCA) method in Section \ref{sec:kpca}. The transformation process is completed in Section \ref{sec:trans_fr}.

\subsection{Kernel Principal Component Analysis}
\label{sec:kpca}

The kPCA method is employed to reduce the dimensionality of the functional predictor. Compared to traditional principal component analysis (PCA), kPCA offers two distinct advantages. Firstly, while traditional PCA is limited to capturing linear relationships between variables, kPCA can capture nonlinear relationships by mapping the original data into a high-dimensional space using a kernel function. This capability allows kPCA to identify intricate structures within the data, enhancing its effectiveness in modelling complex phenomena. Secondly, kPCA provides flexibility in selecting different kernel functions based on the data’s characteristics. By choosing an appropriate kernel, kPCA can effectively represent the data’s underlying features, leading to more accurate and robust results. The utility of kPCA for nonlinear feature extraction has been extensively discussed in the literature \citep[e.g.,][]{scholkopf1998nonlinear, mika1998kernel, rosipal2001kernel, hoffmann2007kernel}.

Consider the matrix of yield curve $Z_t(\cdot)$ at time $t \in \{ 1, \dots, N \} $, evaluated at discrete time points $\tau_1, \tau_2, \dots, \tau_M$: 
\begin{equation*}
    \boldsymbol{Z}_{M \times N} = \left[ \begin{matrix}
    Z_1(\tau_1) & Z_2(\tau_1) & \cdots & Z_N(\tau_1) \\
    Z_1(\tau_2) & Z_2(\tau_2) & \cdots & Z_N(\tau_2) \\
    \vdots & \vdots & \ddots & \vdots \\
    Z_1(\tau_M) & Z_2(\tau_M) & \cdots & Z_N(\tau_M) \\ 
\end{matrix} \right]. 
\end{equation*}
We denote $\boldsymbol{Z}(\tau_i) = \left[ Z_1(\tau_i), Z_2(\tau_i), \dots, Z_N(\tau_i) \right]^\top$ as the time series of the bond yields with maturity time $\tau_i$. Assume $\boldsymbol{\phi}: \mathcal{Y} \to \mathcal{F}$ is a non-linear mapping from the observed input space $\mathcal{Y} \subset \mathbb{R}^N$ to the feature space $\mathcal{F} \subset \mathbb{R}^N$ such that $\boldsymbol{\phi}(\boldsymbol{Z}(\tau_i)) = \left[ \phi_1(\boldsymbol{Z}(\tau_i)), \dots, \phi_N(\boldsymbol{Z}(\tau_i)) \right]^\top$. Let $\boldsymbol{\Phi}$ be an $M \times N$ matrix:
\begin{equation*}
    \boldsymbol{\Phi}_{M \times N} = \left[ \begin{matrix}
    \phi_1(\boldsymbol{Z}(\tau_1)) & \cdots & \phi_N(\boldsymbol{Z}(\tau_1)) \\
    \vdots & \ddots & \vdots \\ 
    \phi_1(\boldsymbol{Z}(\tau_M)) & \cdots & \phi_N(\boldsymbol{Z}(\tau_M)) 
\end{matrix} \right]. 
\end{equation*}
Define $\boldsymbol{B}_{N \times N} = \boldsymbol{\Phi}^\top \boldsymbol{\Phi}$ which is a positive definite matrix. The kernel function $k: \mathcal{Y} \times \mathcal{Y} \to \mathcal{F}$ defines the inner product in the feature space $\mathcal{F}$ and is given by:
\begin{equation}
    k(\boldsymbol{Z}(\tau_i), \boldsymbol{Z}(\tau_j)) = \boldsymbol{\phi}(\boldsymbol{Z}(\tau_i))^\top \boldsymbol{\phi}(\boldsymbol{Z}(\tau_j)) = \sum_{k=1}^N \phi_k(\boldsymbol{Z}(\tau_i)) \phi_k(\boldsymbol{Z}(\tau_j))
    \label{eq:kernel_function}
\end{equation}
for $i, j \in \{ 1, \dots, M \}$. Define $\boldsymbol{K}_{M \times M} = \boldsymbol{\Phi} \boldsymbol{\Phi}^\top$. The objective of kPCA is to find a linear projection that projects $\boldsymbol{\Phi}$ onto uncorrelated components denoted $\boldsymbol{A}$, with lower dimensionality. Each point $\boldsymbol{\phi}(\boldsymbol{Z}(\tau_i))$ can be expressed as a linear combination of $Q \le N$ vectors of dimension $N$: 
\begin{equation*}
    \boldsymbol{\phi}(\boldsymbol{Z}(\tau_i)) = \sum_{q=1}^Q \alpha_{i,q} \boldsymbol{v}_q, 
\end{equation*}
where $\boldsymbol{v}_q$ are orthonormal vectors of dimensions $N$ such that: 
\begin{equation*}
\boldsymbol{v}_q^\top \boldsymbol{v}_k = \sum_{t = 1}^{N} v_{q,t} v_{k,t} = \begin{cases} 
1 \quad \text{if} \; q = k \\ 
0 \quad \text{otherwise} 
\end{cases}
\end{equation*}
Vectors ${\boldsymbol{\alpha}}_q = [ \alpha_{1,q}, \dots, \alpha_{M,q}]^\top$ and ${\boldsymbol{\alpha}}_k = [ \alpha_{1,k}, \dots, \alpha_{M,k}]^\top$ are orthogonal: 
\begin{equation*}
{\boldsymbol{\alpha}}_q^\top {\boldsymbol{\alpha}}_k = \sum_{i=1}^{M} \alpha_{i,q} \alpha_{i,k} = \begin{cases} 
\lambda_q \quad \text{if} \; q = k \\ 
0 \quad \text{otherwise}  
\end{cases} 
\end{equation*}
where $\lambda_q$ is the $q$th eigenvalue of $\boldsymbol{B}$. Define $M \times Q$ matrix $\boldsymbol{A}$ and $Q \times N$ matrix $\boldsymbol{V}$ as:
\begin{equation*}
\boldsymbol{A}_{M \times Q} = [{\boldsymbol{\alpha}}_1, \dots, {\boldsymbol{\alpha}}_Q ] = 
\left[ \begin{matrix} \alpha_{1,1} & \dots & \alpha_{1,Q} \\ 
\vdots & \ddots &  \vdots \\ 
\alpha_{M,1} & \dots & \alpha_{M,Q}  
\end{matrix} \right]_{M \times Q}
\end{equation*}
and
\begin{equation*}
\boldsymbol{V}_{Q \times N} = 
\left[ \begin{matrix} 
\boldsymbol{v}_1^\top \\ 
\vdots \\ 
\boldsymbol{v}_Q^\top
\end{matrix} \right] = 
\left[ \begin{matrix} 
v_{1,1} & \dots & v_{1,N} \\ 
\vdots & \ddots & \vdots \\ 
v_{Q,1} & \dots & v_{Q,N} 
\end{matrix} \right]_{Q \times N}. 
\end{equation*}
Given the assumptions of uncorrelation and orthonormality, we have $\boldsymbol{A}^\top \boldsymbol{A} = \boldsymbol{\Lambda}_{Q \times Q}$ and $\boldsymbol{V} \boldsymbol{V}^\top = \boldsymbol{I}_{Q}$. Therefore, our solution can be written as
\begin{equation}
    \boldsymbol{\Phi}_{M \times N} = \boldsymbol{A}_{M \times Q} \boldsymbol{V}_{Q \times N}. 
\end{equation}

If the non-linear mapping $\boldsymbol{\phi}(\cdot)$ is known, the matrix $\boldsymbol{B}$ is also known and can be rewritten as
\begin{equation*}
    \boldsymbol{B} = \boldsymbol{\Phi}^\top \boldsymbol{\Phi} = \boldsymbol{V}^\top \boldsymbol{A}^\top \boldsymbol{A} \boldsymbol{V} = \boldsymbol{V}^\top \boldsymbol{\Lambda} \boldsymbol{V}. 
\end{equation*}
Therefore, matrix $\boldsymbol{V}$ can be obtained by applying the eigen-decomposition on $\boldsymbol{B}$, and the new representation $\boldsymbol{A}$ of the sample matrix $\boldsymbol{\Phi}$ is obtained by:
\begin{equation}
    \boldsymbol{A} = \boldsymbol{A} \boldsymbol{V} \boldsymbol{V}^\top = \boldsymbol{\Phi} \boldsymbol{V}^\top 
\end{equation}
as the result of the orthonormality of rows in $\boldsymbol{V}$. $\boldsymbol{A}$ is of lower dimension and represents the matrix of the principal components. 

However, the mapping $\boldsymbol{\phi}(\cdot)$ is usually unknown. In this case, the matrix $\boldsymbol{B}$ and the matrix of eigenvectors $\boldsymbol{V}$ are also unknown. One possible solution is given by the employment of the kernel function $k(\cdot, \cdot)$ given in Equation \eqref{eq:kernel_function}. Since $\boldsymbol{B} = \boldsymbol{\Phi}^\top \boldsymbol{\Phi} = \boldsymbol{V}^\top \boldsymbol{\Lambda} \boldsymbol{V}$, we have
\begin{align}
    \boldsymbol{B} &= \boldsymbol{V}^\top \boldsymbol{\Lambda} \boldsymbol{V} \nonumber \\
    \boldsymbol{V} \boldsymbol{B} &= \boldsymbol{V} \boldsymbol{V}^\top \boldsymbol{\Lambda} \boldsymbol{V} \nonumber \\
    \boldsymbol{V} \boldsymbol{\Phi}^\top \boldsymbol{\Phi} &= \boldsymbol{V} \boldsymbol{V}^\top \boldsymbol{\Lambda} \boldsymbol{V} \nonumber \\ 
    \underbrace{ \boldsymbol{V} \boldsymbol{\Phi}^\top }_{=\boldsymbol{A}^\top} \underbrace{ \boldsymbol{\Phi} \boldsymbol{\Phi}^\top }_{=\boldsymbol{K}} &= \underbrace{ \boldsymbol{V} \boldsymbol{V}^\top }_{ = \boldsymbol{I}_{Q}} \boldsymbol{\Lambda} \underbrace{ \boldsymbol{V} \boldsymbol{\Phi}^\top }_{=\boldsymbol{A}^\top} \nonumber
\end{align}
As $\boldsymbol{A} = \boldsymbol{\Phi} \boldsymbol{V}^\top$ and $\boldsymbol{K} = \boldsymbol{\Phi} \boldsymbol{\Phi}^\top$, therefore, by simplifying the last equation, we have
\begin{equation}
    \boldsymbol{A}^\top \boldsymbol{K} = \boldsymbol{\Lambda} \boldsymbol{A}^\top. 
    \label{eq:pc}
\end{equation}
So far, the matrix $\boldsymbol{A}$ is only orthogonal but not orthonormal as $\boldsymbol{A}^\top \boldsymbol{A} = \boldsymbol{\Lambda}$. We define the matrix $\boldsymbol{R}$ as $\boldsymbol{R}_{M \times Q} = \boldsymbol{A} \boldsymbol{\Lambda}^{-\frac{1}{2}}$. Since
\begin{equation*}
    \boldsymbol{R}^\top \boldsymbol{R} = \boldsymbol{\Lambda}^{-\frac{1}{2}} \boldsymbol{A}^\top \boldsymbol{A} \boldsymbol{\Lambda}^{-\frac{1}{2}} = \boldsymbol{\Lambda}^{-\frac{1}{2}} \boldsymbol{\Lambda} \boldsymbol{\Lambda}^{-\frac{1}{2}} = \boldsymbol{I}_Q, 
\end{equation*}
we obtained orthonormal eigenvectors $\boldsymbol{Z}$. By multiplying both sides of Equation \eqref{eq:pc} by $\boldsymbol{\Lambda}^{-\frac{1}{2}}$, we have 
\begin{align}
    \boldsymbol{\Lambda}^{-\frac{1}{2}} \boldsymbol{A}^\top \boldsymbol{K} &= \boldsymbol{\Lambda}^{-\frac{1}{2}} \boldsymbol{\Lambda} \boldsymbol{A}^\top \nonumber \\
    \boldsymbol{R}^\top \boldsymbol{K} &= \boldsymbol{\Lambda} \boldsymbol{R}^\top \nonumber
\end{align}
Therefore, by applying the eigen-decomposition on the matrix $\boldsymbol{K}$, we obtain the matrix $\boldsymbol{R}$ and $\boldsymbol{\Lambda}$, and $\boldsymbol{A}$ is calculated as $\boldsymbol{A} = \boldsymbol{R} \boldsymbol{\Lambda}^{\frac{1}{2}}$. 

Next, we discuss the out-of-sample problem either when the feature mapping $\boldsymbol{\phi}(\cdot)$ is known or unknown. In the first case, since $\boldsymbol{\phi}(\cdot)$ is known, the principal components of the new sample $\boldsymbol{\phi}(\boldsymbol{Z}(\tau^*))$, where $\tau^* \notin \{ \tau_1, \dots, \tau_M \}$ and $\tau^* \in [0, \tau_{max}]$, can be obtained as 
\begin{equation*}
    \boldsymbol{\alpha}^* = \boldsymbol{\phi}(\boldsymbol{Z}(\tau^*)) \boldsymbol{V}^\top. 
\end{equation*}
However, in the second case, we need to define the new observation $\boldsymbol{\phi}(\boldsymbol{Y}(\tau^*))$ in terms of the decomposition of $\boldsymbol{K}$. We have
\begin{align}
    \boldsymbol{\Phi} &= \boldsymbol{A} \boldsymbol{V} \nonumber \\
    \boldsymbol{A}^\top \boldsymbol{\Phi} &= \boldsymbol{A}^\top \boldsymbol{A} \boldsymbol{V} \nonumber \\
    \boldsymbol{A}^\top \boldsymbol{\Phi} &= \boldsymbol{\Lambda} \boldsymbol{V} \nonumber \\
    \boldsymbol{\Lambda}^{-1} \boldsymbol{A}^\top \boldsymbol{\Phi} &= \boldsymbol{V} \nonumber
\end{align}
Define $\boldsymbol{W}_{M \times Q} = \boldsymbol{A} \boldsymbol{\Lambda}^{-1}$. Then, 
\begin{equation*}
    \boldsymbol{v}_q^\top = \sum_{i=1}^M w_{i,q} \boldsymbol{\phi}(\boldsymbol{Z}(\tau_i))
\end{equation*}
and the principal component is given by
\begin{align}
    \alpha_{m,q} &= \boldsymbol{\phi}(\boldsymbol{Z}(\tau_m)) \boldsymbol{v}_q \nonumber \\
    &= \boldsymbol{\phi}(\boldsymbol{Z}(\tau_m)) \sum_{i=1}^N w_{i,q} \boldsymbol{\phi}(\boldsymbol{Z}(\tau_i))^\top \nonumber \\
    &= \sum_{i=1}^M w_{i,q} \boldsymbol{\phi}(\boldsymbol{Z}(\tau_m)) \boldsymbol{\phi}(\boldsymbol{Z}(\tau_i))^\top \nonumber \\
    &= \sum_{i=1}^M w_{i,q} k(\boldsymbol{Z}(\tau_m), \boldsymbol{Z}(\tau_i)). \nonumber
\end{align}
Therefore, given a new sample point $\boldsymbol{Z}(\tau^*)$, its projection can be represented only by the eigen-decomposition of $\boldsymbol{K}$ and the kernel function $k(\cdot, \cdot)$ as 
\begin{equation}
    \alpha_{*, q} = \sum_{i=1}^M w_{i,q} k(\boldsymbol{Z}(\tau^*), \boldsymbol{Z}(\tau_i)). 
\end{equation}

In this paper, we choose the radial basis function (RBF) kernel, which is of the form
\begin{equation}
    k(\boldsymbol{x},\boldsymbol{y}) = \exp{\left(-\frac{||\boldsymbol{x}-\boldsymbol{y}||^2}{2 \sigma^2}\right)}, 
    \label{eq:rbf}
\end{equation}
where $\sigma > 0$ is the hyperparameter. The validation of the RBF kernel is proved in \cite{shawe2004kernel}. Some other choices of kernel functions include polynomial, graph, and ANOVA kernels. 

\subsection{Transformation of Functional Regression}
\label{sec:trans_fr}

In this section, we transform the functional regression component in Equation \eqref{eq:fr_futures} into a weighted sum of finite factors using the Karhunen-Loeve theorem: 

\newtheorem{theorem}{Theorem}

\begin{theorem}[Karhunen-Loeve theorem]
    Suppose $X_t$ is a zero-mean stochastic process for $t \in [a, b]$. $K(s,t)$ is the continuous covariance function. Then $X_t$ can be expressed as 
    \begin{equation*}
        X_t = \sum_{j=1}^\infty Z_j e_j(t),
    \end{equation*}
    where $Z_j = \int_a^b X_t e_j(t) dt$, and $e_j(t)$ are orthonormal basis functions defined in \eqref{eq:orthogonal_basis}. 
\end{theorem}
The orthonormal functions are defined as follows:

\newtheorem{definition}{Definition}

\begin{definition}
    Two real-valued functions $f(x)$ and $g(x)$ are orthonormal over the interval $[a,b]$ if 
    \begin{enumerate}
        \item $\int_a^b f(x) g(x) dx = 0$ 
        \item $||f(x)||_2 = ||g(x)||_2 = \left[ \int_a^b |f(x)|^2 dx \right]^{1/2} = \left[ \int_a^b |g(x)|^2 dx \right]^{1/2} = 1$
    \end{enumerate}
\end{definition}

In this paper, we choose $e_q(\tau_i) = \alpha_{i,q} \boldsymbol{v}_q$ \footnote{In reality, we don't know the eigenvectors $\boldsymbol{v}_q$. Therefore, we choose the eigenvectors of $K$ as a proxy. This will guarantee the orthogonality of the basis functions. } as the orthogonal basis functions, so that $\boldsymbol{\phi} (\boldsymbol{Z}(\tau_i)) = \sum_{q=1}^Q e_q(\tau_i)$. Therefore, we have
\begin{equation}
    e_q(\tau_i) = \alpha_{i,q} \boldsymbol{v}_q = \sum_{j=1}^M w_{j,q} k(\boldsymbol{Z}(\tau_i), \boldsymbol{Z}(\tau_j)) \boldsymbol{v}_q. 
    \label{eq:orthogonal_basis}
\end{equation}

Using the Karhunen-Loeve theorem, we express $Z_t(s)$ and $\gamma_i(s)$ as follows: 
\begin{equation}
    Z_t(s) = \sum_{j=1}^\infty U_{tj} e_j(s)
\end{equation}
and
\begin{equation}
    \gamma_i(s) = \sum_{k=1}^\infty \gamma_{i,k} e_k(s),
\end{equation}
where $U_{tj} = \int_0^{T_M} Z_t(s) e_j(s) ds$ and $\gamma_{i,k} = \int_0^{\tau_M} \gamma_i(s) e_k(s) ds$. 
We then have:
\begin{align}
    \int_0^{\tau_M} \gamma_i(s) Z_t(s) ds =& \int_0^{\tau_M} \left( \sum_{k=1}^\infty \gamma_{i,k} e_k(s) \right) \left( \sum_{j=1}^\infty U_{tj} e_j(s) \right) ds \nonumber \\
    =& \sum_{j=k, j=1}^\infty \gamma_{i,j} U_{tj} \int_0^{\tau_M} \left( e_j(s) \right)^2 ds + \sum_{j \ne k} \gamma_{i,k} U_{tj} \int_0^{\tau_M} e_k(s) e_j(s) ds 
    \nonumber \\
    =& 
    \sum_{j=1}^\infty \gamma_{i,j} U_{tj} \approx \sum_{j=1}^Q \gamma_{i,j} U_{tj}.
\end{align}
Thus, the two-factor functional regression model given in Equations \eqref{eq:fr_futures}, \eqref{eq:fr_chi}, and \eqref{eq:fr_xi} can be rewritten as: 
\begin{equation}
    Y_t(T_i) = A(T_i - t) + e^{-\kappa_{\chi} (T_1 - t)} \chi_t + e^{-\kappa_{\xi} (T_i-t)} \xi_t + \sum_{j=1}^Q \gamma_{i,j} U_{tj} + w_t(T_i), 
\end{equation}
\begin{equation}
    \chi_t = e^{-\kappa_{\chi} \Delta t} \chi_{t-1} + v_{t, \chi}, 
\end{equation}
\begin{equation}
    \xi_t = \frac{\mu_{\xi}}{\kappa_{\xi}} \left(1 - e^{-\kappa_{\xi} \Delta t} \right) + e^{-\kappa_{\xi} \Delta t} \xi_{t-1} + v_{t, \xi}. 
\end{equation}

In matrix notation, this becomes:
\begin{equation}
    \boldsymbol{Y}_t = \boldsymbol{D}_t + \boldsymbol{F}_t \boldsymbol{X}_t + \boldsymbol{\Gamma} \boldsymbol{U}_t + \boldsymbol{w}_t, \;\; \boldsymbol{w}_t \sim N(\boldsymbol{0}, \boldsymbol{\Sigma}_{\boldsymbol{w}}), 
    \label{eq:fr_measurement_matrix}
\end{equation}
\begin{equation}
    \boldsymbol{X}_t = \boldsymbol{C} + \boldsymbol{E} \boldsymbol{X}_{t-1} + \boldsymbol{v}_t, \;\; \boldsymbol{v}_t \sim N(0, \boldsymbol{\Sigma}_{\boldsymbol{v}}). 
    \label{eq:fr_state_matrix}
\end{equation}

\section{Estimation Method}
\label{sec:estimation}

In this section, we discuss the procedures to jointly estimate the latent state variables and unknown model parameters through the Kalman filter. We use the following notations to represent the expectation and covariance matrix of state vector $\boldsymbol{X}_t$: 
\begin{align}
\boldsymbol{a}_{t|t-1} &:= \mathbb{E}(\boldsymbol{X}_t | \boldsymbol{Y}_{1:t-1}),& \boldsymbol{P}_{t|t-1} &:= Cov(\boldsymbol{X}_t | \boldsymbol{Y}_{1:t-1}), \nonumber \\
\boldsymbol{a}_t &:= \mathbb{E}(\boldsymbol{X}_t | \boldsymbol{Y}_{1:t}),& \boldsymbol{P}_t &:= Cov(\boldsymbol{X}_t | \boldsymbol{Y}_{1:t}) \nonumber 
\end{align} 
where $\boldsymbol{Y}_{1:t}$ represents all observations $\boldsymbol{Y}_{1}, \dots, \boldsymbol{Y}_{t}$ up to time $t$.

The system starts from an initial mean vector $\boldsymbol{a}_0$ and an initial covariance $\boldsymbol{P}_0$. Given the estimated state vector $\boldsymbol{a}_{t-1}$ and the covariance matrix $\boldsymbol{P}_{t-1}$ at time $t-1$, we predict the state vector and covariance at time $t$ as
\begin{equation}
    \boldsymbol{a}_{t|t-1} = \boldsymbol{C} + \boldsymbol{E} \boldsymbol{a}_{t-1}
\end{equation}
and 
\begin{equation}
    \boldsymbol{P}_{t|t-1} = \boldsymbol{E} \boldsymbol{P}_{t-1} \boldsymbol{E}^\top + \boldsymbol{\Sigma}_v.
\end{equation}
Then, when new observations $\boldsymbol{Y}_t$ and $\boldsymbol{U}_t$ are available, we update the state vector and covariance as 
\begin{equation}
    \boldsymbol{a}_t = \boldsymbol{a}_{t|t-1} + \boldsymbol{K}_t \left( \boldsymbol{Y}_t - \boldsymbol{D}_t - \boldsymbol{F}_t \boldsymbol{a}_{t|t-1} - \boldsymbol{\Gamma} \boldsymbol{U}_t \right)
\end{equation}
and 
\begin{equation}
    \boldsymbol{P}_t = \left( \boldsymbol{I} - \boldsymbol{K}_t \boldsymbol{F}_t \right) \boldsymbol{P}_{t|t-1},
\end{equation}
where $\boldsymbol{I}$ is the identity matrix and $\boldsymbol{K}_t$ is the Kalman gain matrix given as: 
\begin{equation}
    \boldsymbol{K}_t = \boldsymbol{P}_{t|t-1} \boldsymbol{F}_t^\top \left( \boldsymbol{F}_t \boldsymbol{P}_{t|t-1} \boldsymbol{F}_t^\top + \boldsymbol{\Sigma}_w \right)^{-1}.
\end{equation}
We repeat all the steps for $i \in \{1, \dots, N\}$ to obtain the estimate of the state vector. 

The model parameters, denoted by $\boldsymbol{\theta}$, are estimated using the Maximum Likelihood Estimation (MLE). At time $t$, we define the estimation error of observation $\boldsymbol{Y}_t$ as
\begin{equation}
    \boldsymbol{e}_t = \boldsymbol{Y}_t - \boldsymbol{D}_t - \boldsymbol{F}_t \boldsymbol{a}_{t|t-1} - \boldsymbol{\Gamma} \boldsymbol{U}_t,
\end{equation}
and the covariance is 
\begin{equation}
    \boldsymbol{L}_t = Cov(\boldsymbol{e}_t) = \boldsymbol{F}_t \boldsymbol{P}_{t|t-1} \boldsymbol{F}_t^\top + \boldsymbol{\Sigma}_w. 
\end{equation}
Ignoring the constant terms, the log-likelihood function is given by
\begin{equation}
    l(\boldsymbol{\theta; \boldsymbol{Y}_{1:N}}) = -\frac{1}{2} \sum_{t=1}^N \left( \boldsymbol{e}_t^\top \boldsymbol{L}_t^{-1} \boldsymbol{e}_t + \log{|\boldsymbol{L}_t|} \right).
    \label{eq:ll}
\end{equation}
The set of parameters $\boldsymbol{\theta}$ is estimated by maximising the log-likelihood function \ref{eq:ll}. 

\section{Empirical Analysis}
\label{sec:empirical_analysis}

In this section, we present the results of the empirical analysis using the data introduced in Section \ref{sec:data}. The proposed functional regression model is compared with the traditional Schwartz-Smith two-factor model in Section \ref{sec:result_fr}. In Section \ref{sec:result_st}, we conduct a stress testing analysis by applying two types of shocks, permanent and temporary, to US Treasury yields. We then examine the effects of these shocks on the estimation of futures prices.

\subsection{Data}
\label{sec:data}

\begin{figure}[ht]
    \centering
    \begin{subfigure}{0.45\textwidth}
        \includegraphics[width=\textwidth]{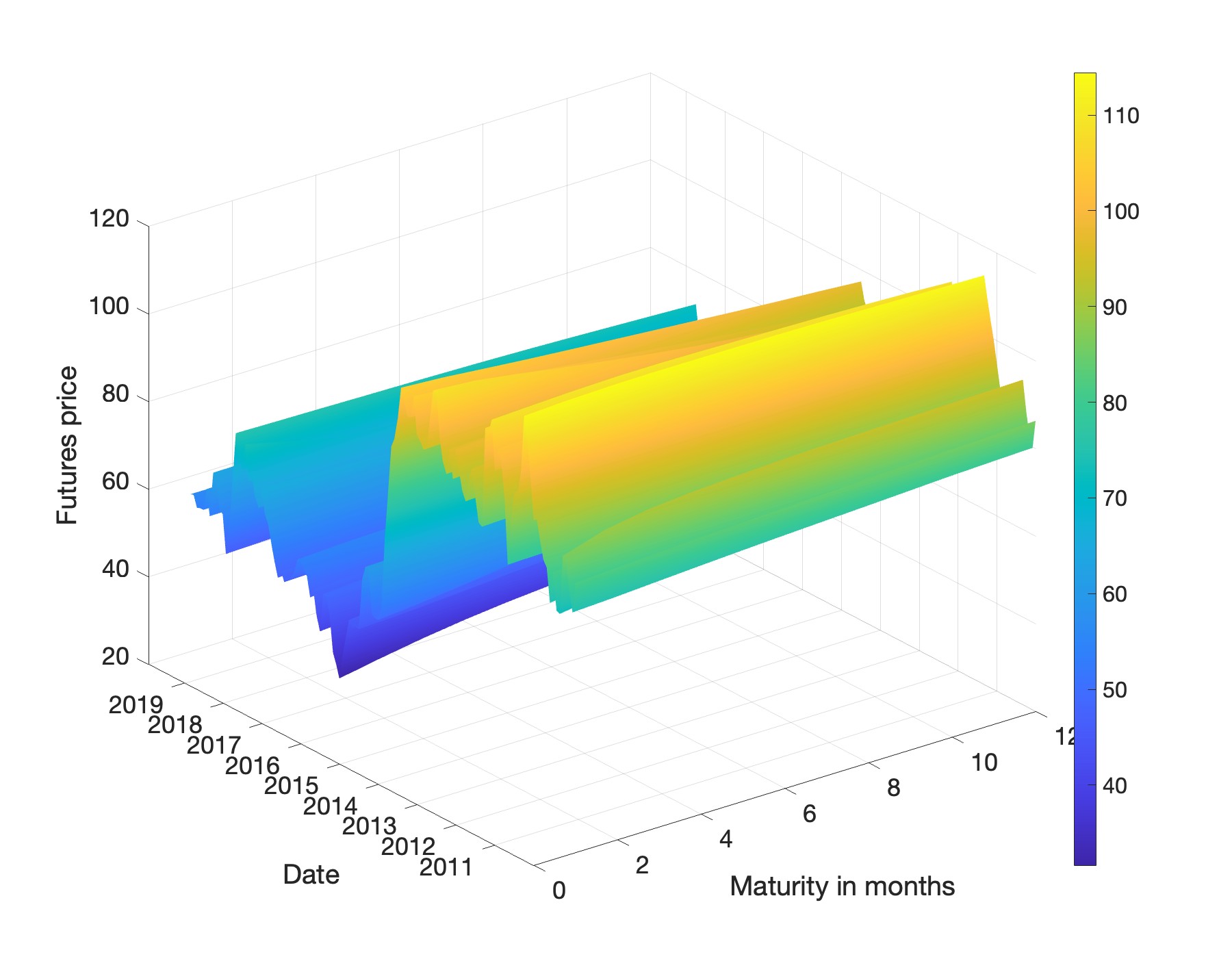}
        \caption{WTI crude oil futures price.}
    \end{subfigure}
    \hfill
    \begin{subfigure}{0.45\textwidth}
        \includegraphics[width=\textwidth]{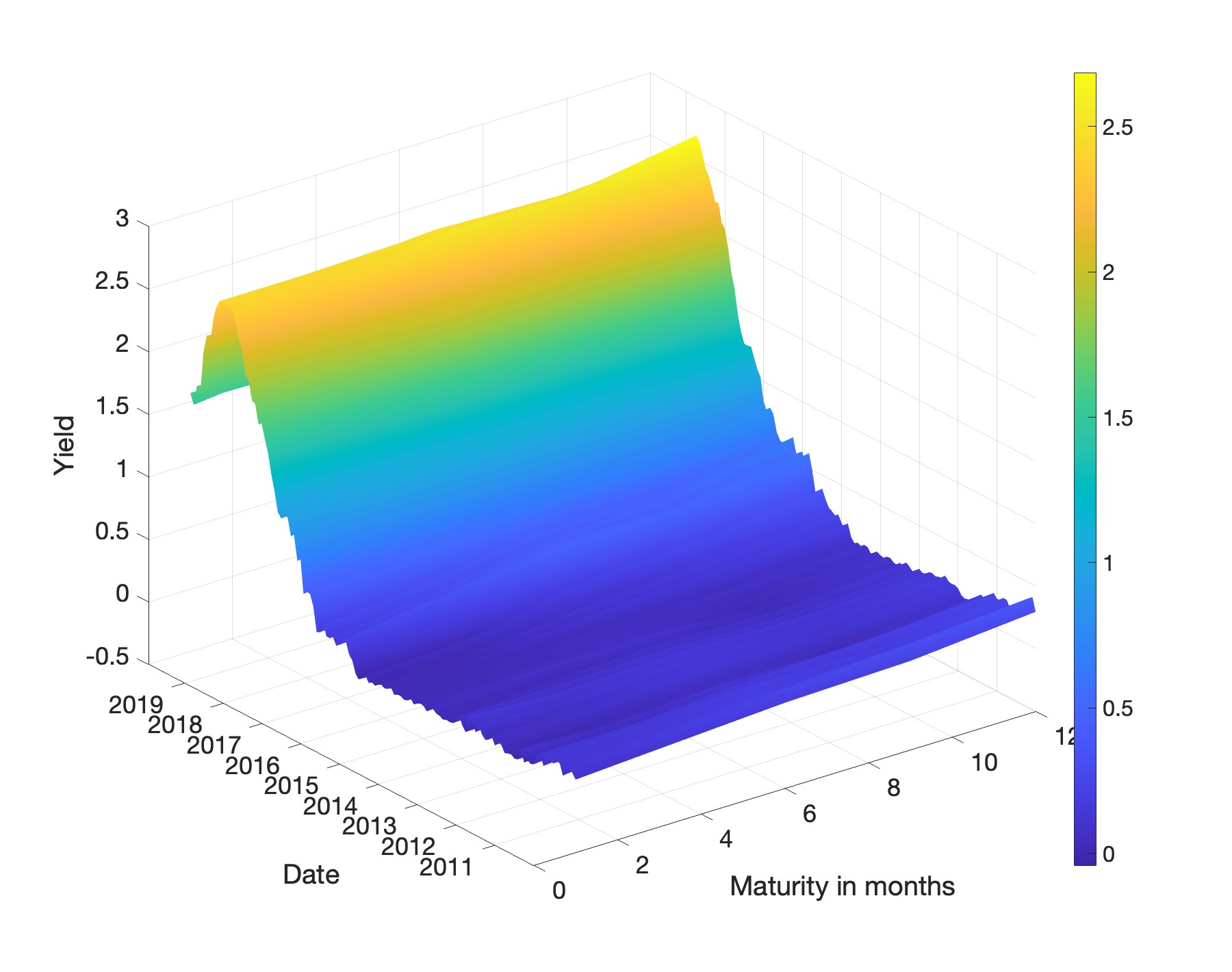}
        \caption{US Treasury yields.}
    \end{subfigure}
    \caption{WTI crude oil futures curve and US Treasury yield curve from 2010 to 2020, both with maturities from 1 to 12 months.}
    \label{fig:data}
\end{figure}

In this paper, we investigate the interdependencies between WTI crude oil futures prices\footnote{The futures prices are quoted on the New York Mercantile Exchange (NYMEX). The data was obtained from LSEG Datascope Select.} and US Treasury yields\footnote{Data was sourced from TradingView.}. We use monthly data for both datasets, covering the period from January 2010 to December 2019. For the futures data, we select contracts with maturities ranging from 1 month to 12 months. The US Treasury yield curve is evaluated at five observed points each month, corresponding to maturities of 1, 3, 6, 9, and 12 months. Figure \ref{fig:data} displays the futures curve (left) and the yield curve (right) for the same period.

\subsection{Two-Factor Functional Regression Model}
\label{sec:result_fr}

In this section, we present the estimation results of the two-factor functional regression model. Throughout the paper, we use the first two factors extracted from US Treasury yields in the functional regression model. The Schwartz-Smith (SS) model serves as the benchmark.

\begin{table}[ht]
    \caption{Root mean square error (RMSE) for each contract using Schwartz-Smith (SS) model and two-factor functional regression (FR) model, respectively.}
    \centering
    \begin{tabular}{ccc}
        \toprule
         Maturity & SS model & FR model \\
         \midrule
         1 month & 1.1631 & 0.9957 \\
         2 months & 0.7021 & 0.5354 \\
         3 months & 0.4038 & 0.2506 \\
         4 months & 0.2144 & 0.0955 \\
         5 months & 0.0827 & 0.0012 \\
         6 months & 0.0035 & 0.0423 \\
         7 months & 0.0373 & 0.0379 \\
         8 months & 0.0374 & 0.0001 \\
         9 months & 0.0001 & 0.0652 \\
         10 months & 0.0658 & 0.1462 \\
         11 months & 0.1484 & 0.2369 \\
         12 months & 0.2413 & 0.3343 \\
         Mean & 0.2583 & 0.2284 \\
         \bottomrule
    \end{tabular}
    \label{tbl:rmse}
\end{table}

Table \ref{tbl:rmse} presents the root mean square error (RMSE) for each futures contract, comparing the SS model and the functional regression model. In general, the functional regression model provides more accurate estimates of the futures curve, with a mean RMSE of 0.2284, compared to 0.2583 for the SS model. The functional regression model outperforms the SS model for short-term contracts, whereas the SS model gives lower RMSEs for long-term contracts.

\begin{figure}[ht]
    \centering
    \includegraphics[width=0.8\textwidth]{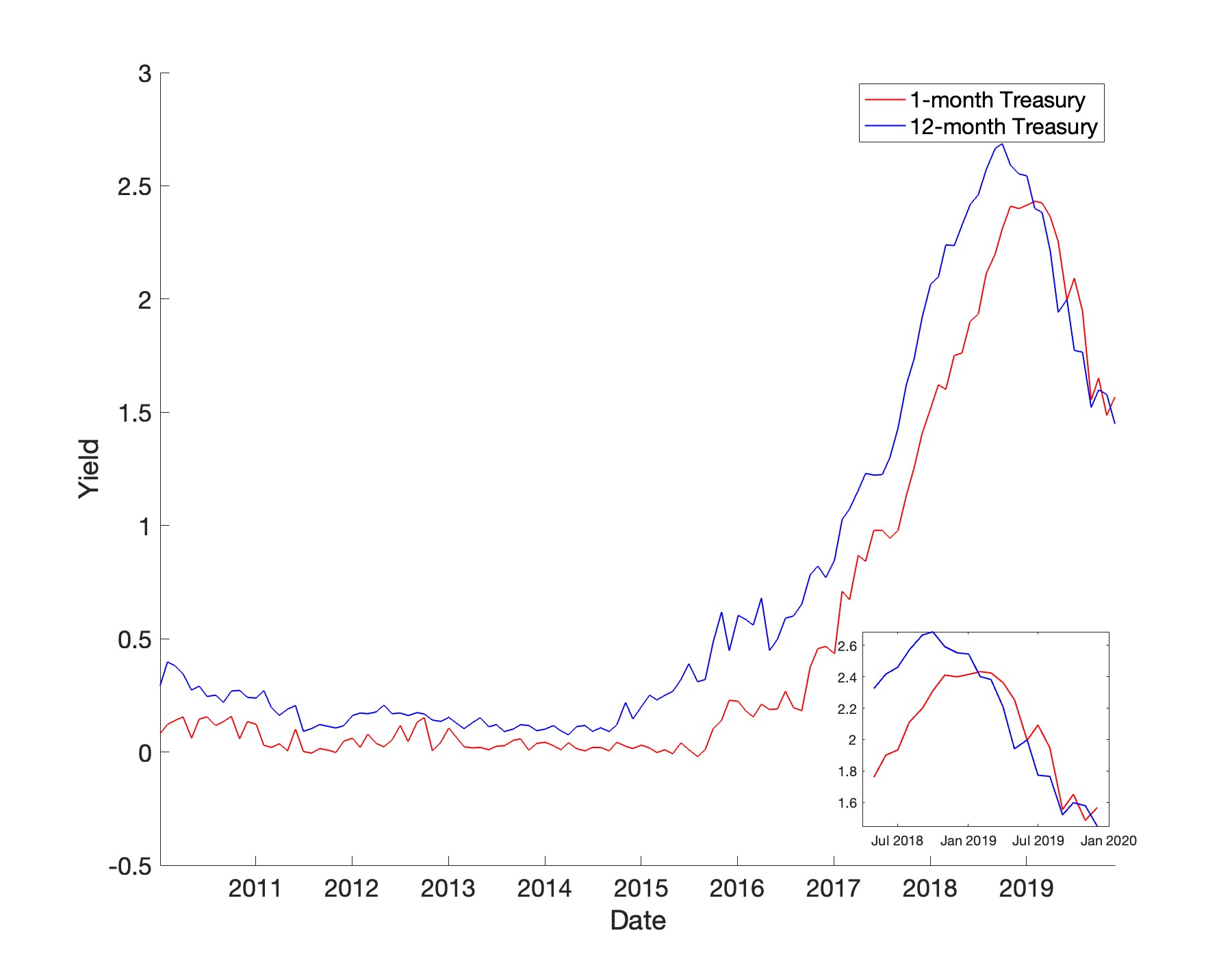}
    \caption{US Treasury yields with maturities 1 month and 12 months. }
    \label{fig:yield}
\end{figure}

Figure \ref{fig:yield} displays US Treasury yields with maturities of 1 month and 12 months. This figure serves as a proxy for identifying periods when the yield curve is in contango, where short-term yields are lower than long-term yields, and when it is in backwardation, where the relationship between short-term and long-term yields is reversed. Furthermore, we use the relative positions of short-term and long-term yields to identify an economic recession. Before January 2019, the 12-month Treasury yield was higher than the 1-month yield, indicating a contango market. However, after January 2019, the yield curve inverted, with the 1-month Treasury yield surpassing the 12-month yield, indicating the beginning of an economic recession.

\begin{figure}[ht]
    \centering
    \includegraphics[width=0.8\textwidth]{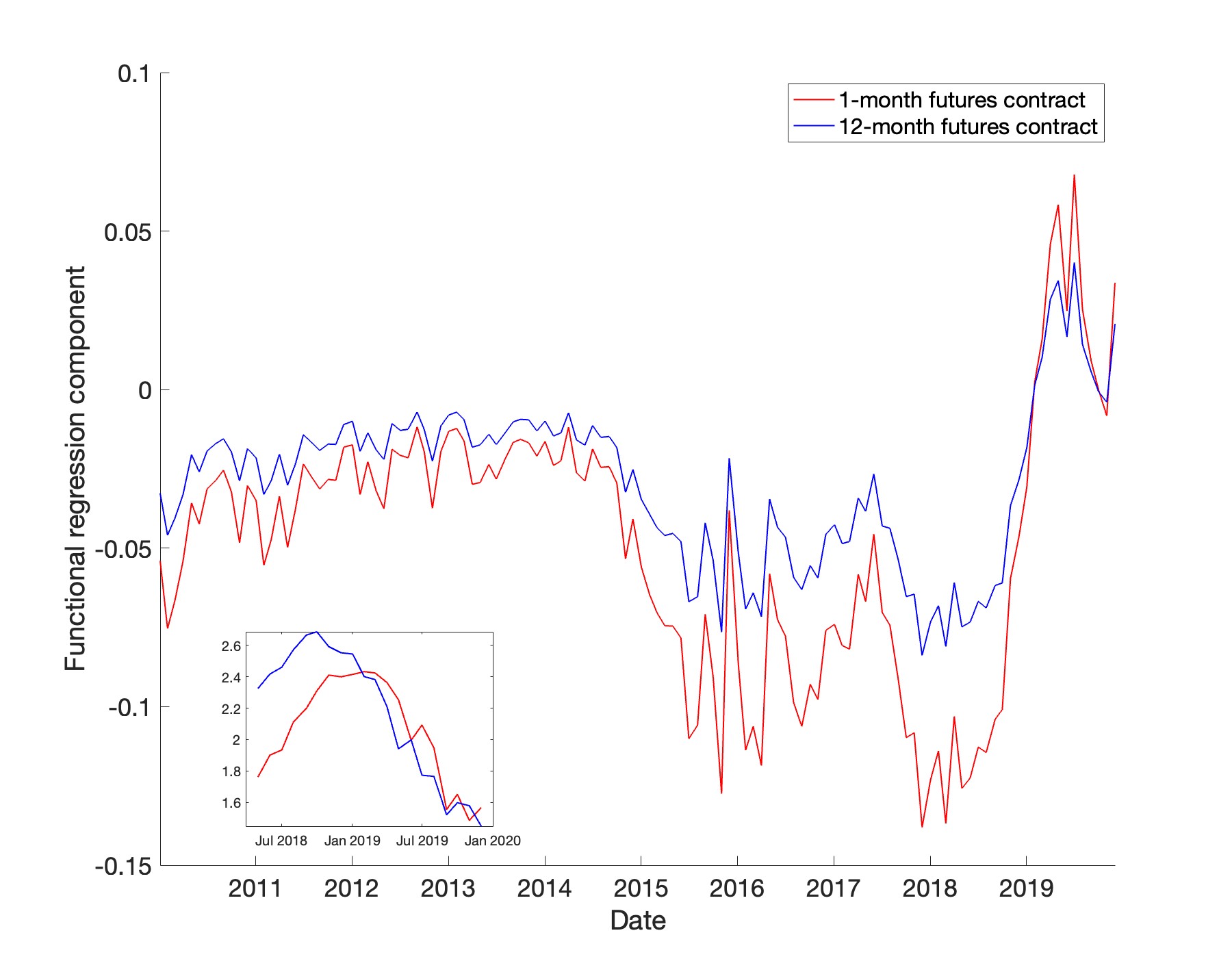}
    \caption{Functional component of the 1-month and 12-month contracts.}
    \label{fig:fr_comp}
\end{figure}

Figure \ref{fig:fr_comp} presents the functional component of the two-factor functional regression model, approximated as:
\begin{equation}
    \int_{0}^{\tau_M} \gamma_i(s) Z_t(s) ds \approx \sum_{j=1}^2 \gamma_{i,j} U_{tj}, 
\end{equation}
for the 1-month and 12-month futures. Before January 2019, when the Treasury market was in contango, the yield curve contributed more significantly to long-term futures, as evidenced by higher values of the functional component. However, after January 2019, during the economic recession, the yield curve had a greater influence on short-term futures than on long-term futures.

For the functional regression model, we estimate the logarithm of the futures price as:
\begin{equation}
    \log{(\hat{F}_{t,T_i})} = A(T_i - t) + e^{-\kappa_{\chi} (T_1 - t)} \chi_t + e^{-\kappa_{\xi} (T_i-t)} \xi_t + \sum_{j=1}^2 \gamma_{i,j} U_{tj},
\end{equation}
Equivalently, the futures price is estimated as:
\begin{equation}
    \hat{F}_{t,T_i} = \exp{\left\{ A(T_i - t) + e^{-\kappa_{\chi} (T_1 - t)} \chi_t + e^{-\kappa_{\xi} (T_i-t)} \xi_t \right\} } \exp{ \left\{ \sum_{j=1}^2 \gamma_{i,j} U_{tj} \right\} }. 
\end{equation}
Before January 2019, the functional component was negative, resulting in a downward adjustment to the SS model. In contrast, after January 2019, the functional component turned positive, leading to an upward adjustment in the SS model.

\begin{figure}[ht]
    \centering
    \includegraphics[width=0.8\textwidth]{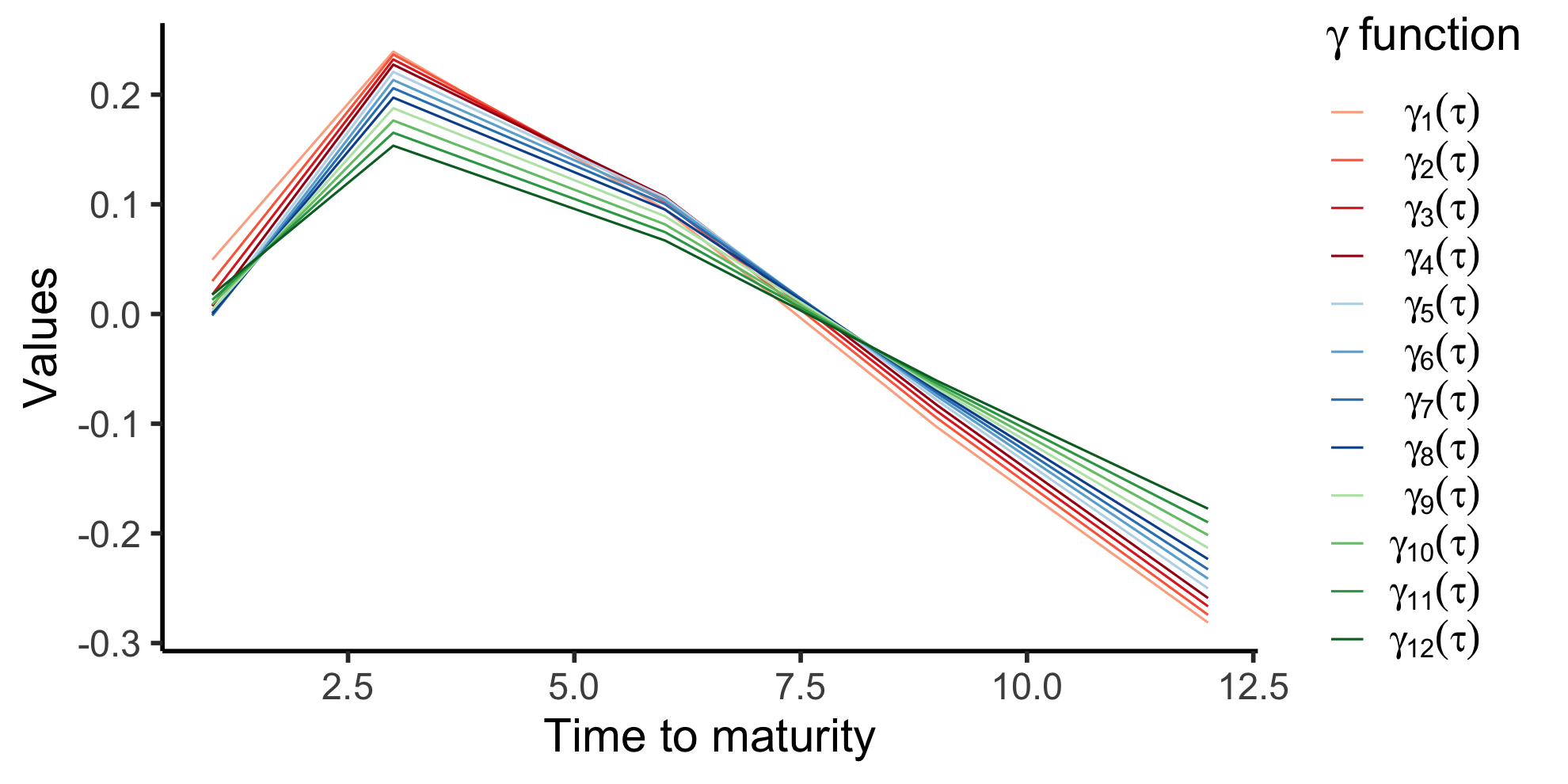}
    \caption{Functional coefficients for WTI crude oil futures with different maturities, using 2 factors extracted from US Treasury yields with maturities less than or equal to 1 year.}
    \label{fig:func_coe}
\end{figure}

Figure \ref{fig:func_coe} gives the estimated functional coefficients $\gamma_i(\tau)$ for $i \in \{ 1, 2, \dots, 12 \}$. The effect of the yield curve on each futures contract follows a similar pattern. Treasuries with maturities close to 3 months exhibit a positive correlation with futures prices, while those with maturities beyond 6 months show a negative correlation.

\FloatBarrier

\subsection{Stress Testing}
\label{sec:result_st}

In this section, we perform a stress testing analysis to examine the impact of shocks applied to Treasury yields on futures prices. Specifically, we aim to address the following research question: How do WTI crude oil futures prices respond to different types of shocks in Treasury yields? To answer this, we define two types of shocks:
\begin{itemize} 
    \item Temporary shock: Between January 2015 and January 2016, all Treasury yields double.
    \item Permanent shock: After January 2015, all Treasury yields double indefinitely.
\end{itemize}
Here, the temporary shock represents short-term disruptions, while the permanent shock reflects long-term structural changes. To conduct this analysis, we first apply these shocks to the US Treasury yields data. Factors are then extracted via kernel principal component analysis (kPCA) using the adjusted data. Finally, we estimate futures prices using the functional regression model described in Equations \eqref{eq:fr_measurement_matrix} and \eqref{eq:fr_state_matrix}.

\begin{figure}[ht]
    \centering
    \begin{subfigure}{0.45\textwidth}
        \includegraphics[width=\textwidth]{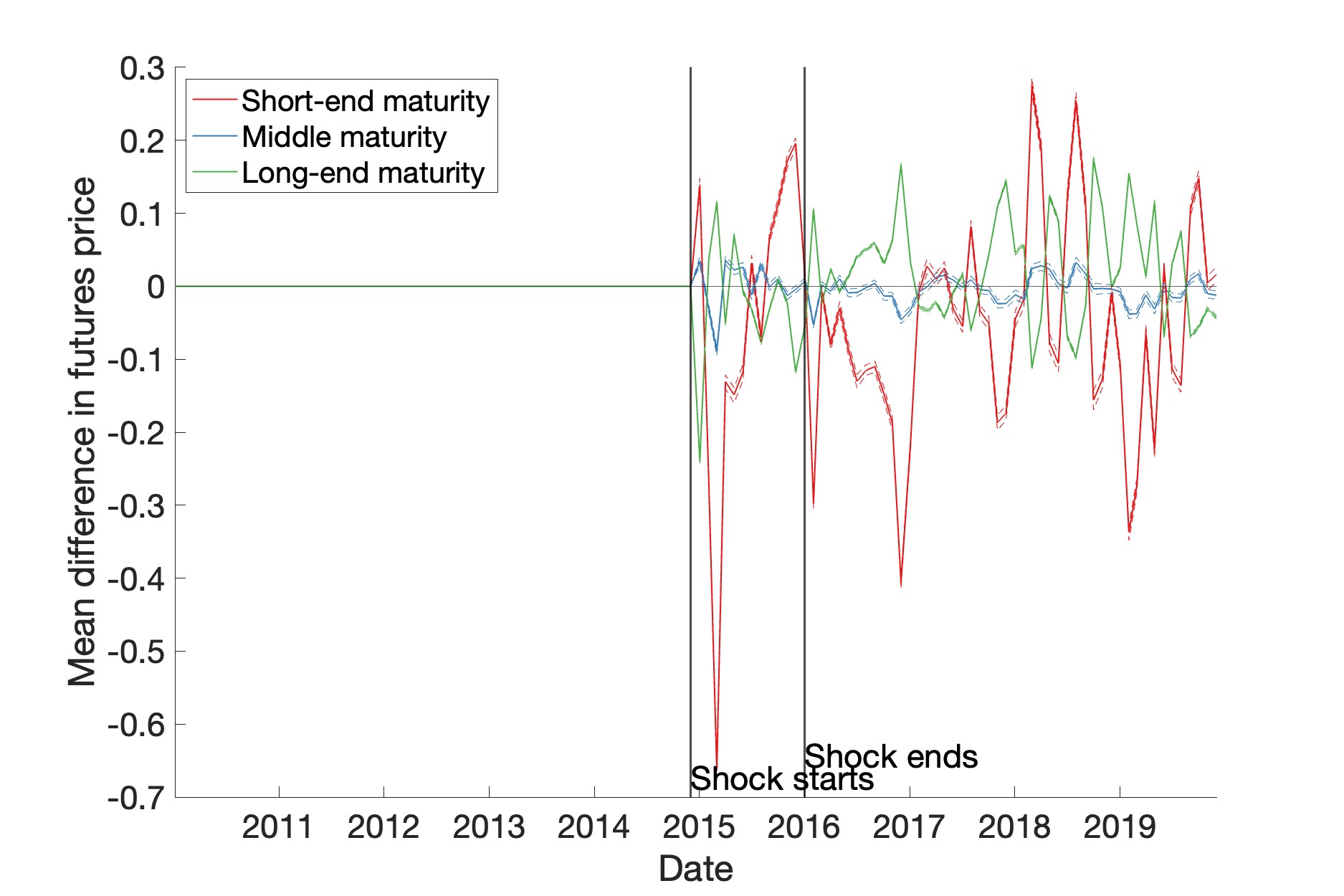}
        \caption{Temporary shock.}
    \end{subfigure}
    \hfill
    \begin{subfigure}{0.45\textwidth}
        \includegraphics[width=\textwidth]{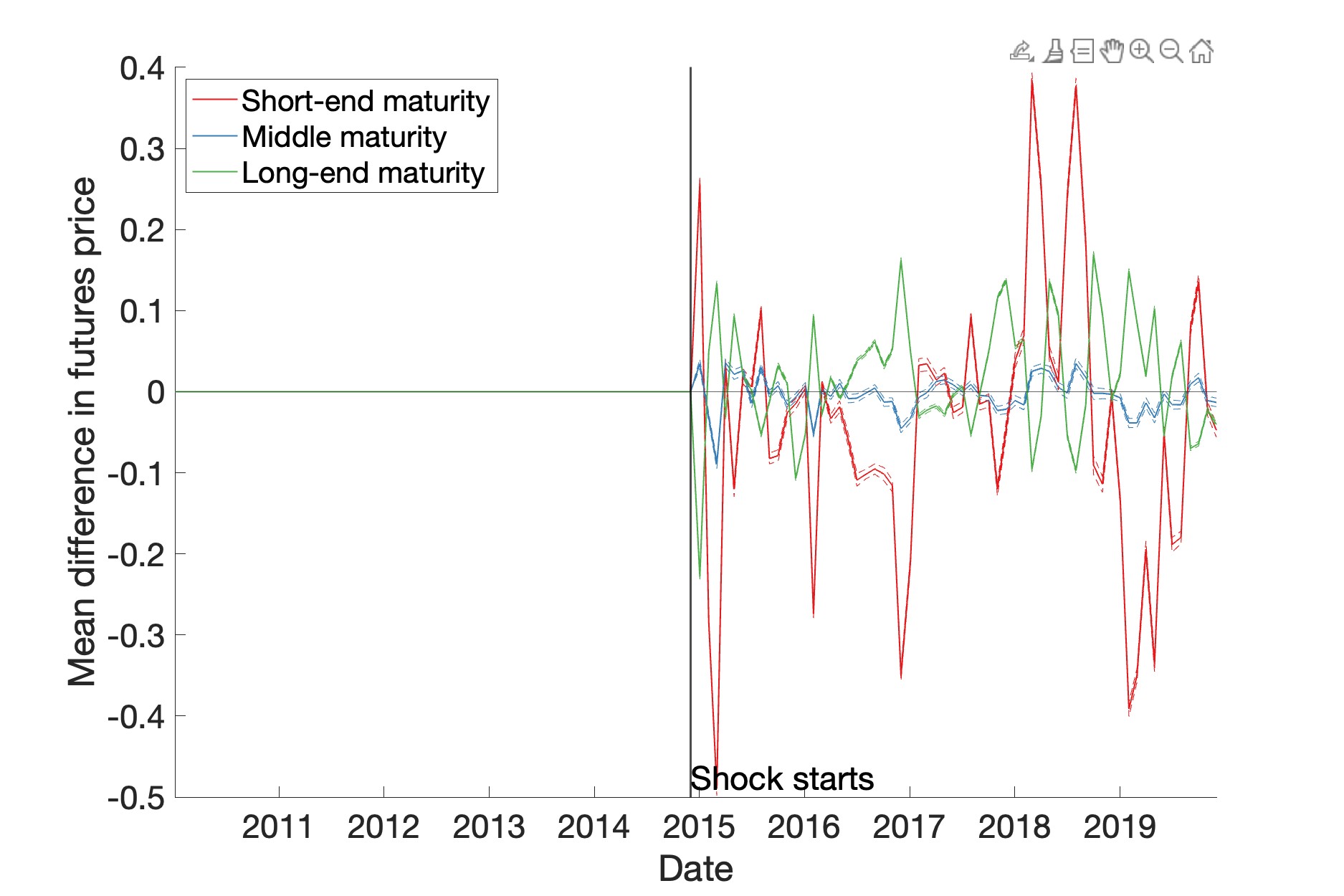}
        \caption{Permanent shock.}
    \end{subfigure}
    \caption{Mean difference (in US dollars) between in-sample WTI crude oil futures price estimations using Treasury data under each shock and original Treasury data. The mean values are taken over short-term ($(0, 4]$ months), middle ($(4, 8]$ months), and long-term ( $(8, 12]$ months) maturities. Dashed lines represent the lower and upper bounds of 95\% confidence interval for each curve.}
    \label{fig:st}
\end{figure}

Figure \ref{fig:st} shows the mean difference in futures price estimates when using the adjusted Treasury data (under each shock) compared to the original Treasury data. We classify the futures contracts into three categories:
\begin{itemize}
    \item Short-term futures: Maturities between 1 and 4 months.
    \item Medium-term futures: Maturities between 5 and 8 months.
    \item Long-term futures: Maturities between 9 and 12 months.
\end{itemize}
The mean values are calculated within each category. In both the temporary and permanent shock scenarios, short-term futures are the most affected, followed by long-term futures. The impact on medium-term futures is minimal. Interestingly, even though the temporary shock ends in January 2016, its effects persist in a long-run period.

\FloatBarrier

\section{Conclusion}
\label{sec:conclusion}

The Schwartz-Smith two-factor model and its extensions have been widely used to estimate commodity futures for over two decades. However, these models are limited by their focus on local factors—those specific to individual markets—without accounting for interdependencies between different markets. In this paper, we proposed a novel two-factor functional regression model that extends the Schwartz-Smith framework by incorporating the interdependencies between the commodity futures market and the bond yields market. Additionally, we applied kernel principal component analysis (kPCA) to transform the functional regression problem into a finite-dimensional estimation problem. The latent short-term and long-term factors, along with the unknown parameters, are estimated jointly using the Kalman filter.

In a comprehensive empirical analysis of WTI crude oil futures, we use US Treasury yields as the functional predictor to explore the relationship between these two markets. We demonstrate that the proposed functional regression model provides more accurate futures price estimates than the Schwartz-Smith model, particularly for short-term contracts. Moreover, we find that under normal economic conditions, when short-term Treasury yields are lower than long-term yields, the original Schwartz-Smith model shrinks after accounting for the effects of Treasury yields, with yields contributing more to long-term futures than short-term futures. In contrast, during an economic recession, indicated by short-term Treasury yields exceeding long-term yields, the Schwartz-Smith model expands, with Treasury yields having a greater influence on short-term futures than on long-term futures.

Furthermore, we conduct a stress testing analysis to assess the impact of two types of shocks, representing short-term disruption and long-term structural change, on the estimated futures prices. Our findings show that both shocks significantly affect short-term futures, with maturities between 1 and 4 months. Notably, the impact of a short-term disruption persists over the long term, even after the shocks end.



\section*{Declaration of competing interests}

The authors have no potential conflicts of interest to disclose.

\section*{Funding}

This research did not receive any specific grant from funding agencies in the public, commercial, or not-for-profit sectors.

\section*{Declaration of generative AI and AI-assisted technologies in the writing process}

While preparing this work, the authors used ChatGPT to improve the manuscript's readability and language. After that, the authors thoroughly reviewed and edited the content, taking full responsibility for the manuscript's content.


\bibliographystyle{elsarticle-harv} 
\bibliography{reference}





\end{document}